\documentclass[10pt,showpacs,preprintnumbers,amsmath,
amssymb,twocolumn,aps,prl,nofootinbib,
superscriptaddress
]{revtex4}

\usepackage{epsfig}
\usepackage{hyperref}
\usepackage{graphicx,epsf}
\usepackage{color}
\usepackage{bm}
\usepackage{psfrag}

\def\be{\begin{equation}}
\def\ee{\end{equation}}
\def\beb{\begin{equation*}}
\def\eeb{\end{equation*}}
\def\bea{\begin{eqnarray}}
\def\eea{\end{eqnarray}}
\def\beab{\begin{eqnarray*}}
\def\eeab{\end{eqnarray*}}

\def\bi{\begin{itemize}}
\def\ei{\end{itemize}}

\def\cs2{c_{\rm{s}}^2}

\def \beg {\begin{enumerate}}
\def \en {\end{enumerate}}

\def\cs{c_{\rm{s}}^2}

\begin{document}
\title{Recombination era magnetic fields from axion dark matter}
\author{Nilanjan Banik}
\email{banik@phys.ufl.edu}
\affiliation{Department of Physics, University of Florida, Gainesville, Florida 32611, USA}
\affiliation{Fermi National Accelerator Laboratory, Batavia, Illinois 60510, USA}
\author{Adam J.~Christopherson}
\email{achristopherson@ufl.edu}
\affiliation{Department of Physics, University of Florida, Gainesville, Florida 32611, USA}

\preprint{FERMILAB-PUB-15-570-A-AE-PPD-T}

\date{\today}

\begin{abstract} 
We introduce a new mechanism for generating magnetic fields in the recombination era. This Harrison-like mechanism utilizes vorticity in baryons that is sourced through the Bose-Einstein condensate of axions via gravitational interactions.  The magnetic fields generated are on galactic scales $\sim 10\,{\rm kpc}$ and have a  magnitude of the order of $B\sim10^{-23}\,{\rm G}$ today. The field has a greater magnitude than those generated from other mechanisms relying on second order perturbation theory, and is sufficient to provide a seed for battery mechanisms.
\end{abstract}


\maketitle


{\bf{\emph{Introduction}} --} 
One of the most pressing problems in modern astronomy is determining the origin of magnetic fields in the Universe. Magnetic fields are observed on all scales, from small scales, such as inside our own solar system, to the largest bound structures, galaxy clusters \cite{Widrow:2002ud, Durrer:2013pga, Giovannini:2003yn}. In fact, recent observations have even detected an intergalactic magnetic field existing in the void regions of the cosmic large scale structure, with magnitude $B_{\rm IGMF}\sim 10^{-18} - 10^{-15}\, {\rm G}$ (e.g., Refs.~\cite{Neronov:1900zz, Tavecchio:2010mk}). Despite their prevalence, there is a large amount of uncertainty in how the first magnetic fields were created.

Setting aside the possibility that magnetic fields were present as initial conditions which is incredibly unappealing, we are yet to fully understand the origin of the first fields. 
These primordial seeds need only be small as there are several ways in which they can then amplified by astrophysical processes; for example, adiabatic contraction or turbulent dynamos during structure formation (see, e.g. Refs.~\cite{Grasso:2000wj, Kulsrud:2007an, ruzmaikin2013magnetic} for a review). The seeds are required to have an amplitude in the range of $10^{-30} - 10^{-20}\,{\rm G}$, with the specific magnitude depending on the details of the dynamo model.  There has been much work over the years addressing the generation of the primordial seed field.

Harrison \cite{harrison,Harrison:1973zz} was one of the first to attempt to explain the origin of the seed magnetic field generated by vorticity in a rotating protogalaxy prior to decoupling. This provided a seed field of the order of $B\sim 10^{-19}\,{\rm G}$, which is large enough to source galactic dynamo mechanisms which enhance this initial seed field to currently observed magnitudes. However, since (at linear order in perturbation theory) vorticity decays \cite{ks}, there is no way to support the vorticity in a post-recombination universe, and so this mechanism of magnetic field generation was criticized \cite{rees:1987}. Later, Mishustin \& Ruzma\u{i}kin \cite{mishustin} investigated the generation of magnetic fields in the post-recombination plasma, finding fields with magnitude of around $10^{-17}\,{\rm G}$ (evaluated at $z\sim100$). Similarly to Harrison, this work required the existence of primordial vorticity. Another recent piece of work used vorticity from the texture scenario of large scale structure formation \cite{sicotte}. However, the resulting field is too weak to act as the primordial seed. The other important seed field mechanism in the early universe is the Biermann battery, either due to shocks \cite{Kulsrud:1996km} or during reionization \cite{1994MNRAS.271L..15S, Gnedin:2000ax}.

Other mechanisms for generating magnetic fields in the very early universe have been studied comprehensively in the literature. These roughly include fields generated during inflation, with a breaking of the conformal invariance of electromagnetism \cite{Turner:1987bw}, during phase-transitions  \cite{Vachaspati:1991nm, Grasso:1997nx} or during (p)reheating \cite{Tornkvist:2000js,Bassett:2000aw}. These all have issues, and sustaining magnetic fields in the very early universe proves to be difficult.

One interesting method for generating magnetic fields around recombination builds upon the ideas of Harrison, and uses second order cosmological perturbation theory \cite{MW2008, MM2008, Christopherson:2011ra}. While vorticity decays at linear order in perturbation theory, there are source terms at second order -- these look very much like the baroclinic term in the Biermann battery -- that allow for vorticity generation \cite{vorticity, Christopherson:2010ek, Christopherson:2010dw}. It is therefore possible that this vorticity comes hand-in-hand with a magnetic field. There have been numerous works to this end \cite{Matarrese:2004kq, Takahashi:2005nd, Gopal:2004ut, Siegel:2006px, Fenu:2010kh, Nalson:2013jya}, that all obtain fields with roughly the same magnitude of $10^{-26}-10^{-23}\,{\rm G}$ at recombination.

In this paper, we consider a new method in which the vorticity can grow around decoupling
by considering axion dark matter.  The axion is a viable dark matter candidate \cite{Abbott:1982af, Preskill:1982cy, Dine:1982ah}, alongside the other major candidate -- the broad class of weakly interacting massive particles (WIMPs). Although axions and WIMPs are similar, the axion is truly a quantum scalar field \cite{Banik:2015ola}, and therefore it is expected that there will be observational differences between the two. In particular, it was recently realized that axion dark matter can form a Bose-Einstein condensate (BEC) \cite{Sikivie:2009qn, Erken:2011dz}. Furthermore, the axion BEC can exhibit vortices, at the zeroes of the wave function \cite{Banik:2013rxa}. It is this vorticity that we will use in the present work to generate  magnetic fields in the early universe.

{\bf{\emph{The axion Bose-Einstein condensate}} --} The axion is motivated by the Peccei-Quinn solution of the strong CP problem, and its mass is thought to be around $10^{-5} \,{\rm eV}/c^2$. Cold axions are one of the leading dark matter candidates. Recently it was shown that axions form a re-thermalizing Bose-Einstein condensate (BEC) through gravitational self-interactions when the photon temperature was around 500 eV \cite{Sikivie:2009qn,Erken:2011dz,Banik:2015sma}. The axion BEC interacts gravitationally with baryons with a relaxation rate 

\be 
\Gamma_{\rm G}\sim4\pi G n m m_{b} \frac{\ell}{\Delta p_{b}}\,,
\ee
where $n$ and  $m$ are the number density and mass of axions respectively, $m_{b}$ is the mass of baryons which is of order 1 GeV, $\ell\sim \frac{1}{H}$ is the correlation length of the axion BEC, $\Delta p_{b}\sim \sqrt{3m_{b}T}$ is the momentum dispersion of the baryons, where $T$ is the photon temperature. 
 
For $T< 1 ~\rm{keV}$ the dominant interaction between photons and baryons is Compton scattering. The relaxation rate for an electron to gain or lose energy by Compton scattering off photons is known from standard cosmology to be \cite{Weinberg:2008zzc}
\begin{align}
\Gamma_{\rm e}\sim 9\times 10^{-21}~\mathrm{s}^{-1}\Omega_{b}h^{2}\left(\frac{T}{T_{\gamma_{0}}}\right)^{4}\,
\end{align}
where $T_{\gamma_{0}}$ is the present day photon temperature and $\Omega_{b}h^{2}$ is the present day physical baryon density parameter. It can be shown that both the ratios $\Gamma_{\rm{G}}/H$ and $\Gamma_{\rm{G}}/\Gamma_{\rm{e}}$ are greater than one around matter radiation equality and keeps increasing thereafter. Therefore the baryons thermalize with the axion BEC. 

{\bf{\emph{Vorticity from tidal torque}} --}
In the standard picture of structure formation baryons collapse onto dark matter overdensities. Tidal torque from nearby inhomogeneities imparts angular momentum onto such protogalaxies. It was shown in Ref.~\cite{Banik:2013rxa} that when a system of axion BEC acquires angular momentum, the axions thermalize and most of them go to a state with minimum $\eta_{i}=\epsilon_{i} - \xi(t) l_{i}$, where $\epsilon_{i}$ and $l_{i}$ are the energy and angular momentum respectively of the $i^{\rm{th}}$ state and $\xi(t)$ is the angular velocity of the system which grows with time as angular momentum grows. This lowest $\eta$ state has non-zero vorticity and therefore the axions acquire a net rotational velocity field.

The baryons being in thermal contact with the axion BEC are dragged along with the axion flow and acquire the same rotational velocity field as the axions. The baryons therefore acquire vorticity from tidal torquing as a result of thermalization with the axion BEC. It should be noted that in general tidal torque on collisionless particles cannot generate rotational flow \cite{Natarajan:2007tk}. Before shell crossing the baryons behave like collisionless particles since dissipative processes and shocks are absent. Therefore if dark matter is made of only WIMPs then there is no vorticity.

%

{\bf{\emph{Magnetic fields}} --} 
We will now show how vorticity in the baryon fluid 
can generate a magnetic field. We follow an approach similar to that of Ref.~\cite{mishustin}. As recombination begins at redshift $z_{\rm r}\sim 1500$, the photon free streaming length grows rapidly such that, on galactic scales, it can be treated as a homogeneous radiation background. The charged particles moving in this radiation background experience a Thomson drag  force, \cite{Peebles:1994xt} $\vec{F}_{\rm T}=\frac{4\sigma_{\rm T}\rho_\gamma}{3c}\vec{v}$, where $\sigma_{\rm T}$ is the Thomson cross-section, $\rho_{\gamma}$ is the energy density of photons and $\vec{v}$ is the velocity of the charged particle relative to the background radiation. Because electrons are much lighter than protons, their acceleration due to this force is much greater than that of protons. Neglecting the effects from neutral species,  the equations of motion for electrons and protons with velocities $\vec{v}_{\rm e}$ and $\vec{v}_{\rm p}$ respectively are 

\begin{align}
\label{eq:v_e}
\frac{d \vec{v_{\rm e}}}{dt}&=-\frac{e}{m_{\rm e}}\Big(\vec{E}+\frac{\vec{v}_{\rm e}}{c}\times\vec{B}\Big)
-\frac{4\sigma_{\rm T}\rho_{\gamma}\vec{v}_{\rm e}}{3cm_{\rm e}}+\frac{\vec{v}_{\rm p}-\vec{v}_{\rm e}}{\tau_{\rm ep}}+\vec{a}_{\rm grav}\,,\\
\label{eq:v_p}
\frac{d \vec{v_{\rm p}}}{dt}&=\frac{e}{m_{\rm p}}\Big(\vec{E}+\frac{\vec{v}_{\rm p}}{c}\times\vec{B}\Big)
-\frac{4\sigma_{\rm T}\rho_{\gamma}\vec{v}_{\rm p}}{3cm_{\rm p}}-\frac{\vec{v}_{\rm p}-\vec{v}_{\rm e}}{\tau_{\rm ep}}+\vec{a}_{\rm grav}\,,
\end{align}
where $\tau_{\rm ep}$ is the characteristic time for momentum exchange via Coulomb scattering, $m_{\rm e}$ and $m_{\rm p}$ are the masses of the electron and proton, respectively, $e$ is the charge of the electron, and $\vec{a}_{\rm grav}$ is the acceleration due to gravitational interactions with the axion BEC. Since we are interested in showing how the vorticity in baryons sourced by the axion BEC can generate magnetic fields, we have neglected the electron and proton pressure terms in the above equations.

The current density is defined as $\vec{J}=en_{\rm e}(\vec{v}_{\rm p}-\vec{v}_{\rm e})$, where $n_{\rm e}\simeq n_{\rm p}$, by local charge neutrality. The Thomson drag term becomes negligibly small after $z \sim 900$ when the timescale of Thomson scattering is greater than the Hubble time. We have assumed that initially there is no magnetic field and neglected the back reaction from the generated magnetic field.

On taking the difference of Eqs.~(\ref{eq:v_e}) and (\ref{eq:v_p}) and neglecting the Thomson drag term for protons, followed by the curl, we arrive at an equation for the vorticity of electron fluid, $\vec{\omega}_{\rm e}$,

\be 
\label{eq:eom}
\frac{d}{dt}\vec{\nabla}\times\Bigg(\frac{\vec{J}}{en_{\rm e}}\Bigg)=\frac{e}{m_{\rm e}}\vec{\nabla}\times\vec{E}+
\frac{4\sigma_{\rm T}\rho_{\gamma}}{3m_{\rm e}c}\vec{\omega}_{\rm e}-2\vec{\nabla}\times\Bigg(\frac{\vec{J}e}{m_{\rm e}\sigma}\Bigg)\,,
\ee
where $\sigma$ is the conductivity of the background medium.
The LHS of Eq.~(\ref{eq:eom}) is negligibly small compared to the first term on the RHS on galactic scales \cite{Widrow:2002ud}. The last term on the RHS of Eq.~(\ref{eq:eom}) is proportional to the magnetic diffusion term which can be neglected because of the high conductivity of the background medium. We are therefore left with
\be 
\label{eq:omega}
\frac{e}{m_{\rm e}}\Big(\vec{\nabla}\times\vec{E}\Big)=-\frac{4\sigma_{\rm T}\rho_{\gamma}}{3m_{\rm e}c}\vec{\omega}_{\rm e}\,.
\ee
On invoking the Maxwell equation
\be 
\frac{1}{c}\frac{\partial \vec{B}}{\partial t}=-\vec{\nabla}\times\vec{E}\,,
\ee
Eq.~(\ref{eq:omega}), becomes
\be 
\label{eq:w}
\frac{\partial \vec{B}}{\partial t}=\frac{4\sigma_{\rm T}\rho_{\gamma}}{3e}\vec{\omega}_{\rm e}\,.
\ee
Of course, these calculations are performed in a static universe, therefore we must transform to the expanding universe in which we live. On doing so, Eq.~(\ref{eq:omega}) becomes
\be 
\label{eq:Bev}
\frac{1}{a^2}\frac{\partial(a^2\vec{B})}{\partial t}=\frac{4\sigma_{\rm T}\rho_{\gamma_0}a^{-4}}{3e}\vec{\omega}_{\rm e}(t)\,,
\ee
where $a(t)$ is the scale factor and a subscript zero denotes the present-day value of a quantity.

Let us consider a galaxy sized ($\sim$ 10 kpc) spherical overdensity of axion BEC onto which baryons are falling. Tidal torque imparts the same specific angular momentum to the infalling matter. Thermalization with the axion BEC results in vorticity in the baryons which is of the order $\omega\sim L/MR^{2}$, where $L$ is the total angular momentum, $M$ is the total mass of the infalling baryons and $R$ is the size of the protogalaxy. Following Peebles \cite{Peebles:1969jm}, the angular momentum of a protogalaxy grows as $t^{5/3}$ in the linear regime, which implies that the vorticity grows as $t^{1/3}$. At $z\sim 10$ the protogalaxies reach their turnaround radius after which they begin to collapse and separate from the background. We denote this redshift by $z_*$ in the following. From this time onwards the evolution is complicated to handle analytically as non linear effects play a significant role. We make an estimate by considering that the angular momentum of the protogalaxy is conserved per comoving volume after they separated from the background, so the vorticity decays like $t^{-4/3}$. 

To summarize in terms of redshift we have
\be
\omega(z) = 
\begin{cases}
\omega_0~(1+z_*)^{5/2} (1+z)^{-1/2}\,, & \, z_*<z<z_{\rm r}\\
\omega_0(1+z)^2\,, & \, 0\leq z \leq z_*\,,
\end{cases}
\ee
where $\omega_0$ is the present day value of the vorticity which, for our galaxy, is $\omega_0\sim10^{-15} ~\rm{s}^{-1}$. Expressing Eq.~(\ref{eq:Bev}) in terms of redshift and using the above expression for vorticity we get an equation which can be integrated from the beginning of recombination upto  $z \sim 900$ when the battery shuts down. For $z_r >z >900$, we have 
\be 
\frac{B(z)}{z^2}\sim 10^{-22}~{\rm G}~\left(\frac{z_{*}}{10}\right)^{5/2}\left(\frac{\omega_{0}}{10^{-15}~\rm{s}^{-1}}\right)
\ln\Big(\frac{z_{\rm r}}{z}\Big)\,.
\ee

The magnetic field grows up to $z\sim 900$ when it has magnitude $B\sim 10^{-17}~\rm{G}$. After this time it is frozen into the residual free charges and decays with the expansion of the universe. The magnetic field today has a magnitude of $B_0\sim10^{-23}~\rm{G}$ on scales of order 10 kpc.  
\\


{\bf{\emph{Discussion}} --} In this paper, we have investigated the generation of magnetic fields from vorticity in the recombination era. We have used a Harrison-like mechanism, with the novelty lying in the fact that the vorticity is not assumed, but rather is inherent in the Bose-Einstein condensate of axions. This provides a natural source of vorticity which is present only for axion dark matter. The magnetic field sourced by this vorticity has a magnitude of $B\sim 10^{-17}\,{\rm G}$ peaking at redshift $z=900$, on scales of $10\, {\rm kpc}$ whose value today is of order $10^{-23}~\rm{G}$. The magnetic field generated through this process acts as a seed for astrophysical amplification mechanisms through the later stages of galaxy formation. There are several different dynamic mechanisms which can amplify seeds by upwards of ten orders of magnitude \cite{Lesch:1994qb, Grasso:2000wj}, and result in the observed fields of the order of a few microGauss at redshift less than one. 

Furthermore, the magnetic field generated from axion dark matter is larger in magnitude that those created by mechanisms relying on higher order fluctuations within the standard $\Lambda$CDM cosmological model. Therefore, this allows for less effective amplification mechanisms to enhance the primordial seed to the observable size.

Finally, we should note that taking into account effects on how the baryons collapse more than the dark matter halo (e.g. Ref.~\cite{1980MNRAS.193..189F}), the field could be diluted by a factor of $(20)^2$ in the inter galactic medium (IGM) compared to the disk. This will result in a field with magnitude $B\sim 10^{-25}{\rm G}$ in the IGM.%
\\


{\bf{\emph{Acknowledgements}} --} 
The authors are grateful to Jim Fry, Karim Malik, Pierre Sikivie, Elisa Todarello, and Richard Woodard for useful comments and discussions. This work is supported in part by the U.S. Department of Energy under Grant No. DE-FG02-97ER41029. Fermilab is operated by Fermi Research Alliance, LLC, under Contract No. DE-AC02-07CH11359 with the U.S. Department of Energy. NB is supported by the Fermilab Graduate Student Research Program in Theoretical Physics.

\bibliography{magfield_axions_v2.bbl}

\begin{thebibliography}{48}
\expandafter\ifx\csname natexlab\endcsname\relax\def\natexlab#1{#1}\fi
\expandafter\ifx\csname bibnamefont\endcsname\relax
  \def\bibnamefont#1{#1}\fi
\expandafter\ifx\csname bibfnamefont\endcsname\relax
  \def\bibfnamefont#1{#1}\fi
\expandafter\ifx\csname citenamefont\endcsname\relax
  \def\citenamefont#1{#1}\fi
\expandafter\ifx\csname url\endcsname\relax
  \def\url#1{\texttt{#1}}\fi
\expandafter\ifx\csname urlprefix\endcsname\relax\def\urlprefix{URL }\fi
\providecommand{\bibinfo}[2]{#2}
\providecommand{\eprint}[2][]{\url{#2}}

\bibitem[{\citenamefont{Widrow}(2002)}]{Widrow:2002ud}
\bibinfo{author}{\bibfnamefont{L.~M.} \bibnamefont{Widrow}},
  \bibinfo{journal}{Rev. Mod. Phys.} \textbf{\bibinfo{volume}{74}},
  \bibinfo{pages}{775} (\bibinfo{year}{2002}), \eprint{astro-ph/0207240}.

\bibitem[{\citenamefont{Durrer and Neronov}(2013)}]{Durrer:2013pga}
\bibinfo{author}{\bibfnamefont{R.}~\bibnamefont{Durrer}} \bibnamefont{and}
  \bibinfo{author}{\bibfnamefont{A.}~\bibnamefont{Neronov}},
  \bibinfo{journal}{Astron.Astrophys.Rev.} \textbf{\bibinfo{volume}{21}},
  \bibinfo{pages}{62} (\bibinfo{year}{2013}), \eprint{1303.7121}.

\bibitem[{\citenamefont{Giovannini}(2004)}]{Giovannini:2003yn}
\bibinfo{author}{\bibfnamefont{M.}~\bibnamefont{Giovannini}},
  \bibinfo{journal}{Int.J.Mod.Phys.} \textbf{\bibinfo{volume}{D13}},
  \bibinfo{pages}{391} (\bibinfo{year}{2004}), \eprint{astro-ph/0312614}.

\bibitem[{\citenamefont{Neronov and Vovk}(2010)}]{Neronov:1900zz}
\bibinfo{author}{\bibfnamefont{A.}~\bibnamefont{Neronov}} \bibnamefont{and}
  \bibinfo{author}{\bibfnamefont{I.}~\bibnamefont{Vovk}},
  \bibinfo{journal}{Science} \textbf{\bibinfo{volume}{328}},
  \bibinfo{pages}{73} (\bibinfo{year}{2010}), \eprint{1006.3504}.

\bibitem[{\citenamefont{Tavecchio et~al.}(2010)\citenamefont{Tavecchio,
  Ghisellini, Foschini, Bonnoli, Ghirlanda et~al.}}]{Tavecchio:2010mk}
\bibinfo{author}{\bibfnamefont{F.}~\bibnamefont{Tavecchio}},
  \bibinfo{author}{\bibfnamefont{G.}~\bibnamefont{Ghisellini}},
  \bibinfo{author}{\bibfnamefont{L.}~\bibnamefont{Foschini}},
  \bibinfo{author}{\bibfnamefont{G.}~\bibnamefont{Bonnoli}},
  \bibinfo{author}{\bibfnamefont{G.}~\bibnamefont{Ghirlanda}},
  \bibnamefont{et~al.}, \bibinfo{journal}{Mon.Not.Roy.Astron.Soc.}
  \textbf{\bibinfo{volume}{406}}, \bibinfo{pages}{L70} (\bibinfo{year}{2010}),
  \eprint{1004.1329}.

\bibitem[{\citenamefont{Grasso and Rubinstein}(2001)}]{Grasso:2000wj}
\bibinfo{author}{\bibfnamefont{D.}~\bibnamefont{Grasso}} \bibnamefont{and}
  \bibinfo{author}{\bibfnamefont{H.~R.} \bibnamefont{Rubinstein}},
  \bibinfo{journal}{Phys.Rept.} \textbf{\bibinfo{volume}{348}},
  \bibinfo{pages}{163} (\bibinfo{year}{2001}), \eprint{astro-ph/0009061}.

\bibitem[{\citenamefont{Kulsrud and Zweibel}(2008)}]{Kulsrud:2007an}
\bibinfo{author}{\bibfnamefont{R.~M.} \bibnamefont{Kulsrud}} \bibnamefont{and}
  \bibinfo{author}{\bibfnamefont{E.~G.} \bibnamefont{Zweibel}},
  \bibinfo{journal}{Rept. Prog. Phys.} \textbf{\bibinfo{volume}{71}},
  \bibinfo{pages}{0046091} (\bibinfo{year}{2008}), \eprint{0707.2783}.

\bibitem[{\citenamefont{Ruzmaikin et~al.}(2013)\citenamefont{Ruzmaikin,
  Sokoloff, and Shukurov}}]{ruzmaikin2013magnetic}
\bibinfo{author}{\bibfnamefont{A.}~\bibnamefont{Ruzmaikin}},
  \bibinfo{author}{\bibfnamefont{D.}~\bibnamefont{Sokoloff}}, \bibnamefont{and}
  \bibinfo{author}{\bibfnamefont{A.}~\bibnamefont{Shukurov}},
  \emph{\bibinfo{title}{Magnetic Fields of Galaxies}}, Astrophysics and Space
  Science Library (\bibinfo{publisher}{Springer Netherlands},
  \bibinfo{year}{2013}), ISBN \bibinfo{isbn}{9789400928350},
  \urlprefix\url{https://books.google.com/books?id=ZZz-CAAAQBAJ}.

\bibitem[{\citenamefont{Harrison}(1970)}]{harrison}
\bibinfo{author}{\bibfnamefont{E.~R.} \bibnamefont{Harrison}},
  \bibinfo{journal}{Mon. Not. R. atr. Soc.} \textbf{\bibinfo{volume}{147}},
  \bibinfo{pages}{279} (\bibinfo{year}{1970}).

\bibitem[{\citenamefont{Harrison}(1973)}]{Harrison:1973zz}
\bibinfo{author}{\bibfnamefont{E.}~\bibnamefont{Harrison}},
  \bibinfo{journal}{Phys.Rev.Lett.} \textbf{\bibinfo{volume}{30}},
  \bibinfo{pages}{188} (\bibinfo{year}{1973}).

\bibitem[{\citenamefont{Kodama and Sasaki}(1984)}]{ks}
\bibinfo{author}{\bibfnamefont{H.}~\bibnamefont{Kodama}} \bibnamefont{and}
  \bibinfo{author}{\bibfnamefont{M.}~\bibnamefont{Sasaki}},
  \bibinfo{journal}{Prog. Theor. Phys. Suppl.} \textbf{\bibinfo{volume}{78}},
  \bibinfo{pages}{1} (\bibinfo{year}{1984}).

\bibitem[{\citenamefont{{Rees}}(1987)}]{rees:1987}
\bibinfo{author}{\bibfnamefont{M.~J.} \bibnamefont{{Rees}}},
  \bibinfo{journal}{Quarterly Journal of the Royal Astronomical Society}
  \textbf{\bibinfo{volume}{28}}, \bibinfo{pages}{197} (\bibinfo{year}{1987}).

\bibitem[{\citenamefont{Mishustin and Ruzmaikin}(1972)}]{mishustin}
\bibinfo{author}{\bibfnamefont{I.~N.} \bibnamefont{Mishustin}}
  \bibnamefont{and} \bibinfo{author}{\bibfnamefont{A.~A.}
  \bibnamefont{Ruzmaikin}}, \bibinfo{journal}{Journal of Experimental and
  Theoretical Physics} \textbf{\bibinfo{volume}{61}}, \bibinfo{pages}{441}
  (\bibinfo{year}{1972}).

\bibitem[{\citenamefont{{Sicotte}}(1997)}]{sicotte}
\bibinfo{author}{\bibfnamefont{H.}~\bibnamefont{{Sicotte}}},
  \bibinfo{journal}{Mon.Not.Roy.Astron.Soc.} \textbf{\bibinfo{volume}{287}},
  \bibinfo{pages}{1} (\bibinfo{year}{1997}).

\bibitem[{\citenamefont{Kulsrud et~al.}(1997)\citenamefont{Kulsrud, Cen,
  Ostriker, and Ryu}}]{Kulsrud:1996km}
\bibinfo{author}{\bibfnamefont{R.~M.} \bibnamefont{Kulsrud}},
  \bibinfo{author}{\bibfnamefont{R.}~\bibnamefont{Cen}},
  \bibinfo{author}{\bibfnamefont{J.~P.} \bibnamefont{Ostriker}},
  \bibnamefont{and} \bibinfo{author}{\bibfnamefont{D.}~\bibnamefont{Ryu}},
  \bibinfo{journal}{Astrophys. J.} \textbf{\bibinfo{volume}{480}},
  \bibinfo{pages}{481} (\bibinfo{year}{1997}), \eprint{astro-ph/9607141}.

\bibitem[{\citenamefont{{Subramanian} et~al.}(1994)\citenamefont{{Subramanian},
  {Narasimha}, and {Chitre}}}]{1994MNRAS.271L..15S}
\bibinfo{author}{\bibfnamefont{K.}~\bibnamefont{{Subramanian}}},
  \bibinfo{author}{\bibfnamefont{D.}~\bibnamefont{{Narasimha}}},
  \bibnamefont{and} \bibinfo{author}{\bibfnamefont{S.~M.}
  \bibnamefont{{Chitre}}}, \bibinfo{journal}{Mon.Not.Roy.Astron.Soc.}
  \textbf{\bibinfo{volume}{271}}, \bibinfo{pages}{L15} (\bibinfo{year}{1994}).

\bibitem[{\citenamefont{Gnedin et~al.}(2000)\citenamefont{Gnedin, Ferrara, and
  Zweibel}}]{Gnedin:2000ax}
\bibinfo{author}{\bibfnamefont{N.~Y.} \bibnamefont{Gnedin}},
  \bibinfo{author}{\bibfnamefont{A.}~\bibnamefont{Ferrara}}, \bibnamefont{and}
  \bibinfo{author}{\bibfnamefont{E.~G.} \bibnamefont{Zweibel}},
  \bibinfo{journal}{Astrophys. J.} \textbf{\bibinfo{volume}{539}},
  \bibinfo{pages}{505} (\bibinfo{year}{2000}), \eprint{astro-ph/0001066}.

\bibitem[{\citenamefont{Turner and Widrow}(1988)}]{Turner:1987bw}
\bibinfo{author}{\bibfnamefont{M.~S.} \bibnamefont{Turner}} \bibnamefont{and}
  \bibinfo{author}{\bibfnamefont{L.~M.} \bibnamefont{Widrow}},
  \bibinfo{journal}{Phys.Rev.} \textbf{\bibinfo{volume}{D37}},
  \bibinfo{pages}{2743} (\bibinfo{year}{1988}).

\bibitem[{\citenamefont{Vachaspati}(1991)}]{Vachaspati:1991nm}
\bibinfo{author}{\bibfnamefont{T.}~\bibnamefont{Vachaspati}},
  \bibinfo{journal}{Phys.Lett.} \textbf{\bibinfo{volume}{B265}},
  \bibinfo{pages}{258} (\bibinfo{year}{1991}).

\bibitem[{\citenamefont{Grasso and Riotto}(1998)}]{Grasso:1997nx}
\bibinfo{author}{\bibfnamefont{D.}~\bibnamefont{Grasso}} \bibnamefont{and}
  \bibinfo{author}{\bibfnamefont{A.}~\bibnamefont{Riotto}},
  \bibinfo{journal}{Phys.Lett.} \textbf{\bibinfo{volume}{B418}},
  \bibinfo{pages}{258} (\bibinfo{year}{1998}), \eprint{hep-ph/9707265}.

\bibitem[{\citenamefont{Tornkvist et~al.}(2001)\citenamefont{Tornkvist, Davis,
  Dimopoulos, and Prokopec}}]{Tornkvist:2000js}
\bibinfo{author}{\bibfnamefont{O.}~\bibnamefont{Tornkvist}},
  \bibinfo{author}{\bibfnamefont{A.-C.} \bibnamefont{Davis}},
  \bibinfo{author}{\bibfnamefont{K.}~\bibnamefont{Dimopoulos}},
  \bibnamefont{and} \bibinfo{author}{\bibfnamefont{T.}~\bibnamefont{Prokopec}},
  \bibinfo{journal}{AIP Conf.Proc.} \textbf{\bibinfo{volume}{555}},
  \bibinfo{pages}{443} (\bibinfo{year}{2001}), \eprint{astro-ph/0011278}.

\bibitem[{\citenamefont{Bassett et~al.}(2001)\citenamefont{Bassett, Pollifrone,
  Tsujikawa, and Viniegra}}]{Bassett:2000aw}
\bibinfo{author}{\bibfnamefont{B.~A.} \bibnamefont{Bassett}},
  \bibinfo{author}{\bibfnamefont{G.}~\bibnamefont{Pollifrone}},
  \bibinfo{author}{\bibfnamefont{S.}~\bibnamefont{Tsujikawa}},
  \bibnamefont{and} \bibinfo{author}{\bibfnamefont{F.}~\bibnamefont{Viniegra}},
  \bibinfo{journal}{Phys.Rev.} \textbf{\bibinfo{volume}{D63}},
  \bibinfo{pages}{103515} (\bibinfo{year}{2001}), \eprint{astro-ph/0010628}.

\bibitem[{\citenamefont{Malik and Wands}(2009)}]{MW2008}
\bibinfo{author}{\bibfnamefont{K.~A.} \bibnamefont{Malik}} \bibnamefont{and}
  \bibinfo{author}{\bibfnamefont{D.}~\bibnamefont{Wands}},
  \bibinfo{journal}{Phys. Rept.} \textbf{\bibinfo{volume}{475}},
  \bibinfo{pages}{1} (\bibinfo{year}{2009}), \eprint{0809.4944}.

\bibitem[{\citenamefont{Malik and Matravers}(2008)}]{MM2008}
\bibinfo{author}{\bibfnamefont{K.~A.} \bibnamefont{Malik}} \bibnamefont{and}
  \bibinfo{author}{\bibfnamefont{D.~R.} \bibnamefont{Matravers}},
  \bibinfo{journal}{Class. Quant. Grav.} \textbf{\bibinfo{volume}{25}},
  \bibinfo{pages}{193001} (\bibinfo{year}{2008}), \eprint{0804.3276}.

\bibitem[{\citenamefont{Christopherson}(2011)}]{Christopherson:2011ra}
\bibinfo{author}{\bibfnamefont{A.~J.} \bibnamefont{Christopherson}}
  (\bibinfo{year}{2011}), \eprint{1106.0446}.

\bibitem[{\citenamefont{Christopherson
  et~al.}(2009)\citenamefont{Christopherson, Malik, and Matravers}}]{vorticity}
\bibinfo{author}{\bibfnamefont{A.~J.} \bibnamefont{Christopherson}},
  \bibinfo{author}{\bibfnamefont{K.~A.} \bibnamefont{Malik}}, \bibnamefont{and}
  \bibinfo{author}{\bibfnamefont{D.~R.} \bibnamefont{Matravers}},
  \bibinfo{journal}{Phys. Rev.} \textbf{\bibinfo{volume}{D79}},
  \bibinfo{pages}{123523} (\bibinfo{year}{2009}), \eprint{0904.0940}.

\bibitem[{\citenamefont{Christopherson
  et~al.}(2011)\citenamefont{Christopherson, Malik, and
  Matravers}}]{Christopherson:2010ek}
\bibinfo{author}{\bibfnamefont{A.~J.} \bibnamefont{Christopherson}},
  \bibinfo{author}{\bibfnamefont{K.~A.} \bibnamefont{Malik}}, \bibnamefont{and}
  \bibinfo{author}{\bibfnamefont{D.~R.} \bibnamefont{Matravers}},
  \bibinfo{journal}{Phys.Rev.} \textbf{\bibinfo{volume}{D83}},
  \bibinfo{pages}{123512} (\bibinfo{year}{2011}), \eprint{1008.4866}.

\bibitem[{\citenamefont{Christopherson and
  Malik}(2011)}]{Christopherson:2010dw}
\bibinfo{author}{\bibfnamefont{A.~J.} \bibnamefont{Christopherson}}
  \bibnamefont{and} \bibinfo{author}{\bibfnamefont{K.~A.} \bibnamefont{Malik}},
  \bibinfo{journal}{Class.Quant.Grav.} \textbf{\bibinfo{volume}{28}},
  \bibinfo{pages}{114004} (\bibinfo{year}{2011}), \eprint{1010.4885}.

\bibitem[{\citenamefont{Matarrese et~al.}(2005)\citenamefont{Matarrese,
  Mollerach, Notari, and Riotto}}]{Matarrese:2004kq}
\bibinfo{author}{\bibfnamefont{S.}~\bibnamefont{Matarrese}},
  \bibinfo{author}{\bibfnamefont{S.}~\bibnamefont{Mollerach}},
  \bibinfo{author}{\bibfnamefont{A.}~\bibnamefont{Notari}}, \bibnamefont{and}
  \bibinfo{author}{\bibfnamefont{A.}~\bibnamefont{Riotto}},
  \bibinfo{journal}{Phys. Rev.} \textbf{\bibinfo{volume}{D71}},
  \bibinfo{pages}{043502} (\bibinfo{year}{2005}), \eprint{astro-ph/0410687}.

\bibitem[{\citenamefont{Takahashi et~al.}(2005)\citenamefont{Takahashi, Ichiki,
  Ohno, and Hanayama}}]{Takahashi:2005nd}
\bibinfo{author}{\bibfnamefont{K.}~\bibnamefont{Takahashi}},
  \bibinfo{author}{\bibfnamefont{K.}~\bibnamefont{Ichiki}},
  \bibinfo{author}{\bibfnamefont{H.}~\bibnamefont{Ohno}}, \bibnamefont{and}
  \bibinfo{author}{\bibfnamefont{H.}~\bibnamefont{Hanayama}},
  \bibinfo{journal}{Phys. Rev. Lett.} \textbf{\bibinfo{volume}{95}},
  \bibinfo{pages}{121301} (\bibinfo{year}{2005}), \eprint{astro-ph/0502283}.

\bibitem[{\citenamefont{Gopal and Sethi}(2005)}]{Gopal:2004ut}
\bibinfo{author}{\bibfnamefont{R.}~\bibnamefont{Gopal}} \bibnamefont{and}
  \bibinfo{author}{\bibfnamefont{S.}~\bibnamefont{Sethi}},
  \bibinfo{journal}{Mon. Not. Roy. Astron. Soc.}
  \textbf{\bibinfo{volume}{363}}, \bibinfo{pages}{521} (\bibinfo{year}{2005}),
  \eprint{astro-ph/0411170}.

\bibitem[{\citenamefont{Siegel and Fry}(2006)}]{Siegel:2006px}
\bibinfo{author}{\bibfnamefont{E.~R.} \bibnamefont{Siegel}} \bibnamefont{and}
  \bibinfo{author}{\bibfnamefont{J.~N.} \bibnamefont{Fry}},
  \bibinfo{journal}{Astrophys. J.} \textbf{\bibinfo{volume}{651}},
  \bibinfo{pages}{627} (\bibinfo{year}{2006}), \eprint{astro-ph/0604526}.

\bibitem[{\citenamefont{Fenu et~al.}(2011)\citenamefont{Fenu, Pitrou, and
  Maartens}}]{Fenu:2010kh}
\bibinfo{author}{\bibfnamefont{E.}~\bibnamefont{Fenu}},
  \bibinfo{author}{\bibfnamefont{C.}~\bibnamefont{Pitrou}}, \bibnamefont{and}
  \bibinfo{author}{\bibfnamefont{R.}~\bibnamefont{Maartens}},
  \bibinfo{journal}{Mon.Not.Roy.Astron.Soc.} \textbf{\bibinfo{volume}{414}},
  \bibinfo{pages}{2354} (\bibinfo{year}{2011}), \eprint{1012.2958}.

\bibitem[{\citenamefont{Nalson et~al.}(2014)\citenamefont{Nalson,
  Christopherson, and Malik}}]{Nalson:2013jya}
\bibinfo{author}{\bibfnamefont{E.}~\bibnamefont{Nalson}},
  \bibinfo{author}{\bibfnamefont{A.~J.} \bibnamefont{Christopherson}},
  \bibnamefont{and} \bibinfo{author}{\bibfnamefont{K.~A.} \bibnamefont{Malik}},
  \bibinfo{journal}{JCAP} \textbf{\bibinfo{volume}{1409}}, \bibinfo{pages}{023}
  (\bibinfo{year}{2014}), \eprint{1312.6504}.

\bibitem[{\citenamefont{Abbott and Sikivie}(1983)}]{Abbott:1982af}
\bibinfo{author}{\bibfnamefont{L.}~\bibnamefont{Abbott}} \bibnamefont{and}
  \bibinfo{author}{\bibfnamefont{P.}~\bibnamefont{Sikivie}},
  \bibinfo{journal}{Phys.Lett.} \textbf{\bibinfo{volume}{B120}},
  \bibinfo{pages}{133} (\bibinfo{year}{1983}).

\bibitem[{\citenamefont{Preskill et~al.}(1983)\citenamefont{Preskill, Wise, and
  Wilczek}}]{Preskill:1982cy}
\bibinfo{author}{\bibfnamefont{J.}~\bibnamefont{Preskill}},
  \bibinfo{author}{\bibfnamefont{M.~B.} \bibnamefont{Wise}}, \bibnamefont{and}
  \bibinfo{author}{\bibfnamefont{F.}~\bibnamefont{Wilczek}},
  \bibinfo{journal}{Phys. Lett.} \textbf{\bibinfo{volume}{B120}},
  \bibinfo{pages}{127} (\bibinfo{year}{1983}).

\bibitem[{\citenamefont{Dine and Fischler}(1983)}]{Dine:1982ah}
\bibinfo{author}{\bibfnamefont{M.}~\bibnamefont{Dine}} \bibnamefont{and}
  \bibinfo{author}{\bibfnamefont{W.}~\bibnamefont{Fischler}},
  \bibinfo{journal}{Phys. Lett.} \textbf{\bibinfo{volume}{B120}},
  \bibinfo{pages}{137} (\bibinfo{year}{1983}).

\bibitem[{\citenamefont{Banik et~al.}(2015)\citenamefont{Banik, Christopherson,
  Sikivie, and Todarello}}]{Banik:2015ola}
\bibinfo{author}{\bibfnamefont{N.}~\bibnamefont{Banik}},
  \bibinfo{author}{\bibfnamefont{A.~J.} \bibnamefont{Christopherson}},
  \bibinfo{author}{\bibfnamefont{P.}~\bibnamefont{Sikivie}}, \bibnamefont{and}
  \bibinfo{author}{\bibfnamefont{E.~M.} \bibnamefont{Todarello}},
  \bibinfo{journal}{Phys. Rev. D} \textbf{\bibinfo{volume}{91}},
  \bibinfo{pages}{123540} (\bibinfo{year}{2015}), \eprint{1504.05968}.

\bibitem[{\citenamefont{Sikivie and Yang}(2009)}]{Sikivie:2009qn}
\bibinfo{author}{\bibfnamefont{P.}~\bibnamefont{Sikivie}} \bibnamefont{and}
  \bibinfo{author}{\bibfnamefont{Q.}~\bibnamefont{Yang}},
  \bibinfo{journal}{Phys.Rev.Lett.} \textbf{\bibinfo{volume}{103}},
  \bibinfo{pages}{111301} (\bibinfo{year}{2009}), \eprint{0901.1106}.

\bibitem[{\citenamefont{Erken et~al.}(2012)\citenamefont{Erken, Sikivie, Tam,
  and Yang}}]{Erken:2011dz}
\bibinfo{author}{\bibfnamefont{O.}~\bibnamefont{Erken}},
  \bibinfo{author}{\bibfnamefont{P.}~\bibnamefont{Sikivie}},
  \bibinfo{author}{\bibfnamefont{H.}~\bibnamefont{Tam}}, \bibnamefont{and}
  \bibinfo{author}{\bibfnamefont{Q.}~\bibnamefont{Yang}},
  \bibinfo{journal}{Phys.Rev.} \textbf{\bibinfo{volume}{D85}},
  \bibinfo{pages}{063520} (\bibinfo{year}{2012}), \eprint{1111.1157}.

\bibitem[{\citenamefont{Banik and Sikivie}(2013)}]{Banik:2013rxa}
\bibinfo{author}{\bibfnamefont{N.}~\bibnamefont{Banik}} \bibnamefont{and}
  \bibinfo{author}{\bibfnamefont{P.}~\bibnamefont{Sikivie}},
  \bibinfo{journal}{Phys.Rev.} \textbf{\bibinfo{volume}{D88}},
  \bibinfo{pages}{123517} (\bibinfo{year}{2013}), \eprint{1307.3547}.

\bibitem[{\citenamefont{Banik and Sikivie}(2015)}]{Banik:2015sma}
\bibinfo{author}{\bibfnamefont{N.}~\bibnamefont{Banik}} \bibnamefont{and}
  \bibinfo{author}{\bibfnamefont{P.}~\bibnamefont{Sikivie}}
  (\bibinfo{year}{2015}), \eprint{1501.05913}.

\bibitem[{\citenamefont{Weinberg}(2008)}]{Weinberg:2008zzc}
\bibinfo{author}{\bibfnamefont{S.}~\bibnamefont{Weinberg}}
  (\bibinfo{year}{2008}).

\bibitem[{\citenamefont{Natarajan and Sikivie}(2008)}]{Natarajan:2007tk}
\bibinfo{author}{\bibfnamefont{A.}~\bibnamefont{Natarajan}} \bibnamefont{and}
  \bibinfo{author}{\bibfnamefont{P.}~\bibnamefont{Sikivie}},
  \bibinfo{journal}{Phys.Rev.} \textbf{\bibinfo{volume}{D77}},
  \bibinfo{pages}{043531} (\bibinfo{year}{2008}), \eprint{0711.1297}.

\bibitem[{\citenamefont{Peebles}(1994)}]{Peebles:1994xt}
\bibinfo{author}{\bibfnamefont{P.}~\bibnamefont{Peebles}}
  (\bibinfo{year}{1994}).

\bibitem[{\citenamefont{Peebles}(1969)}]{Peebles:1969jm}
\bibinfo{author}{\bibfnamefont{P.}~\bibnamefont{Peebles}},
  \bibinfo{journal}{Astrophys.J.} \textbf{\bibinfo{volume}{155}},
  \bibinfo{pages}{393} (\bibinfo{year}{1969}).

\bibitem[{\citenamefont{Lesch and Chiba}(1995)}]{Lesch:1994qb}
\bibinfo{author}{\bibfnamefont{H.}~\bibnamefont{Lesch}} \bibnamefont{and}
  \bibinfo{author}{\bibfnamefont{M.}~\bibnamefont{Chiba}},
  \bibinfo{journal}{Astron.Astrophys.} \textbf{\bibinfo{volume}{297}},
  \bibinfo{pages}{305} (\bibinfo{year}{1995}), \eprint{astro-ph/9411072}.

\bibitem[{\citenamefont{{Fall} and {Efstathiou}}(1980)}]{1980MNRAS.193..189F}
\bibinfo{author}{\bibfnamefont{S.~M.} \bibnamefont{{Fall}}} \bibnamefont{and}
  \bibinfo{author}{\bibfnamefont{G.}~\bibnamefont{{Efstathiou}}},
  \bibinfo{journal}{Mon.Not.Roy.Astron.Soc.} \textbf{\bibinfo{volume}{193}},
  \bibinfo{pages}{189} (\bibinfo{year}{1980}).

\end{thebibliography}

\end{document}